\documentclass[conference,letterpaper]{IEEEtran}   	
\IEEEoverridecommandlockouts 

\usepackage[utf8]{inputenc}
\usepackage{amsmath,amssymb,amsfonts,amsthm}
\usepackage[numbers,sort&compress]{natbib}

\usepackage{bm}
\usepackage{color}
\usepackage{glossaries}
\usepackage{graphicx}
\usepackage{multirow}
\usepackage[linesnumbered]{algorithm2e}
\usepackage{algpseudocode}
\usepackage[colorlinks]{hyperref}

\newacronym{FA}{FA}{fluid antenna}
\newacronym{FAMA}{FAMA}{fluid antenna multiple access}
\newacronym{FAS}{FAS}{fluid antenna system}
\newacronym{LoS}{LoS}{line-of-sight}
\newacronym{MIMO}{MIMO}{multiple-input multiple-output}
\newacronym{NLoS}{NLoS}{non-line-of-sight}
\newacronym{RF}{RF}{radio-frequency}
\newacronym{SINR}{SINR}{signal-to-interference-plus-noise ratio}
\newacronym{SNR}{SNR}{signal-to-noise ratio}
\newacronym{SIW}{SIW}{substrate-integrated waveguide}

\renewcommand{\vec}[1]{\mathbf{\lowercase{#1}}}	   
\newcommand{\mat}[1]{\mathbf{\uppercase{#1}}}	   
\newcommand{\real}[1]{\text{Re}\left\{#1\right\}}	

\definecolor{dkgreen}{rgb}{0.4, 0.7, 0.2}

\newtheorem{remark}{Remark}

\title{Implementing Fluid Antennas in the Beamspace: Performance Evaluation and Codebook Design\\
\thanks{The work of P. Ramírez-Espinosa has been funded by the European Union under the Marie Sklodowska-Curie grant agreement No. 101109529. The work of F.J. López-Martínez and D. Morales-Jimenez is supported in part by the State Research Agency (AEI) of Spain and the European Social Fund under grant RYC2020-030536-I and by MICIU/AEI/10.13039/501100011033 and FEDER/UE under grant PID2023-149975OB-I00 (COSTUME).}
}

\author{Pablo Ram\'irez-Espinosa\IEEEauthorrefmark{1}, F. Javier López-Martínez\IEEEauthorrefmark{2} and David Morales-Jim\'enez\IEEEauthorrefmark{2}\\
\IEEEauthorblockA{\IEEEauthorrefmark{1}Telecommunications Research Institute (TELMA), University of M\'alaga, M\'alaga 29071 (Spain)}
\IEEEauthorblockA{\IEEEauthorrefmark{2}Department of Signal Theory, Networking and Communications, University of Granada, Granada 18071 (Spain)\\
Email: pre@ic.uma.es, fjlm@ugr.es,  dmorales@ugr.es
}}

\begin{document}
\maketitle

\begin{abstract}
    Metasurface-based fluid antenna systems (FASs) have been recently proposed as an inexpensive, scalable and practical alternative implementation for the fluid-antenna concept. This work thoroughly evaluates the performance of metasurface-based FASs in the context of multi-user communications. We extend the state-of-the-art signal model of FASs to electronically-reconfigurable designs, explicitly including the antenna response in the equivalent channel and resulting correlation structure. A general codebook design procedure, accounting for practical aspects like reflections and radiation efficiency, is presented and used to design the different antenna configurations (regarded as FAS ports). Importantly, we show that, with proper design, metasurface-based FASs can significantly outperform conceptual ones. While state-of-the-art theoretical embodiments of FAS rely on spatial flexibility for constructive/destructive interference, metasurface-based FASs exploit interference cancellation through projection onto the interference null space. Numerical results show a remarkable improvement when the system is dominated by interference (i.e., the natural FASs operational regime), regardless of spatial propagation characteristics.     
\end{abstract}

\section{Introduction}
\label{sec:intro}
\IEEEPARstart{T}{he} \gls{FA} concept has evolved from alternative and flexible (in shape and/or radiation properties) antenna implementations \cite{Dey2016, Hayes2012} to an extreme spatial diversity system where a single antenna element can be freely moved within a predefined aperture \cite{Wong2022_BruceLee}. With \gls{FAMA} as the chief application to enable multi-user communications in an open-loop fashion \cite{Wong2022}, \glspl{FAS} have recently become one of the most actively researched topics in the wireless community. While this interest has led to a large number of publications, many practical questions such as: 1) how FASs can be implemented, and 2) whether the widely used signal model for \glspl{FAS} is valid for realistic implementations, still remain largely unanswered. 

The original motivation behind \glspl{FAS} is that the antenna can be positioned at arbitrary points, ideally sampling the channel at any location in the space continuum, such that system performance is fundamentally bounded by the channel spatial correlation \cite{Khammassi2023, Wong2022, Ramirez2024}. In practice, there are different alternatives to physically implement the concept of \gls{FA}. For instance, mechanical-based designs (either moving the antenna itself or some liquid metal) have been proposed to materialize conceptual \gls{FAS} \cite{Zhu2024, Shen2024}, although they have been deemed impractical due to limitations in accuracy, switching speed, scalability, and power consumption. 

Electronically reconfigurable \glspl{FAS}, where physical movement is replaced by radiation pattern reconfiguration---thus effectively sampling different (yet correlated) baseband channels at the \gls{RF} chain---have been considered as a more promising alternative. Two different implementations, both sharing the same working principle, have been proposed to date: \textit{i)} pixel-based \cite{Zhang2024,Zhang2025, Liu2026}, and \textit{ii)} metasurface-based \cite{Ramirez2025,Liu2025,Zhang2026}. These implementations efficiently materialize the concept of \glspl{FAS} without the drawbacks of mechanical implementations, but lead to the second question posed above. Given that they replace \textit{physical} spatial positioning with movement along the antenna \textit{beamspace}, it is legitimate to ask: \textit{what is the impact of this operational principle on the signal model and the inherent channel correlation structure?} This evidently raises a follow-up question: \textit{are all the conclusions in the large body of published works applicable to electronically reconfigurable \glspl{FAS}?} 

Regarding the former question, since reconfiguration is done in the beamspace, then all the electromagnetic characteristics of the antenna must be accounted for in both the signal model and correlation structure. Also, as illustrated in \cite{Zhang2024, Ramirez2025}, the resulting ``spatial" correlation matrix now depends on the available antenna configurations. This leads to the problem of correlation design and, more explicitly, codebook\footnote{The term `codebook' here represents the collection of available (designed) antenna configurations in electronically reconfigurable (metasurface-based) FASs, not to be confused with precoding codebooks in conventional MIMO.} design for electronically reconfigurable \glspl{FAS} \cite{Zhang2026}. \glspl{FAS} performance no longer depends solely on antenna positioning and channel spatial correlation (imposed by the geometry of the propagation scenario), as assumed in most existing works.  

Motivated by the above research gap, this paper investigates the communications aspects of metasurface-based \glspl{FAS}. Inheriting the electromagnetically compliant signal model in \cite{Ramirez2025}, we propose a generic codebook generation procedure and evaluate its impact on the resulting correlation structure. Besides, we evaluate the performance of metasurface-based \glspl{FAS} in terms of multiplexing capabilities and achievable rate, showing that electronically reconfigurable implementations \textit{can significantly outperform} conceptual \glspl{FAS}.

\textit{Notation:} Vectors and matrices are represented by bold lowercase and uppercase symbols, respectively. $(\cdot)^*$, $(\cdot)^T$ and $(\cdot)^H$ denote conjugate, transpose and conjugate transpose. $\mathbb{E}_{x}[\cdot]$ is the mathematical expectation over $x$, $\real{\cdot}$ denote real part, $i = \sqrt{-1}$ is the imaginary number, and $\|\cdot\|_2$ is the $\ell_2$ norm of a vector. Any other specific notation will be defined when necessary. 

\textit{Supplementary files: } the \textsc{Matlab} code used for simulations and figures in this paper can be found at:\\ \url{https://github.com/preugr/Metasurface-based-FAS.git}

\section{Signal model for metasurface-based FASs}
\label{sec:signal_model}
\subsection{General received signal and covariance structure}

Consider as a starting point the downlink signal model widely used in conceptual \glspl{FAS} known as slow FAMA \cite{Wong2022}, where symbols for $M\geq 1$ users are directly transmitted (no precoding involved) by $M$ different classical (fixed-position) antennas\footnote{If the transmitter uses more antennas than users, then redundant antennas can be seen as additional channel scatterers since no precoding is applied.}. Each user is equipped with a fluid antenna with $L$ ports available. Therefore, the received signal at user $m$ when selecting port $l$ is given by:
\begin{equation}
    s_l^{(m)} = h_l^{(m,m)}x_m + \sum_{\widetilde{m}\neq m}^M h_l^{(\widetilde{m},m)}x_{\widetilde{m}} + w_l^{(m)}, \label{eq:FAS_signal}
\end{equation}
where $x_m$ is the symbol intended for user $m$, $w_l^{(m)}$ is the noise term, and $h_l^{(\widetilde{m},m)}$ is the channel from transmitting antenna $\widetilde{m}$ to the $l$-th port of user $m$. Stacking all channel samples for user $m$ in a vector $\vec{h}^{(\widetilde{m},m)} = \begin{pmatrix}h_l^{(\widetilde{m},m)} & \dots & h_L^{(\widetilde{m},m)}\end{pmatrix}^T$, then $\vec{h}^{(\widetilde{m},m)}$ is obtained by sampling the spatial channel at positions dictated by the $L$ ports \cite{Ramirez2024}. 

In electronically reconfigurable \glspl{FAS}, the antenna positions are fixed, and reconfigurability is achieved by varying the radiation pattern. Consequently, the $L$ ports no longer represent arbitrary positions where the spatial channel may be sampled, but achievable antenna configurations (equivalently, radiation patterns). Hence, they can be seen as \textit{virtual} ports where, instead of in space, the channel is sampled in the \textit{beamspace}. The signal model in \eqref{eq:FAS_signal} is still valid though, as long as $\vec{h}^{(\widetilde{m},m)}$ represents the equivalent channel seen by the \gls{RF} chain, accounting therefore for the antenna characteristics. In general, we can express it as
\begin{equation}
    \vec{h}^{(\widetilde{m},m)} = \mat{G}^T\vec{f}^{(\widetilde{m},m)},
\end{equation}
where $\vec{f}^{(\widetilde{m},m)}\in\mathbb{C}^{N\times 1}$ is the wireless channel from transmitting antenna $\widetilde{m}$ to the $N$ antenna elements composing the electronically reconfigurable \gls{FAS} of user $m$, and 
\begin{equation}
    \mat{G}\in\mathbb{C}^{N\times L} = \begin{pmatrix} \vec{g}_1 & \dots & \vec{g}_l & \dots & \vec{g}_L\end{pmatrix}
\end{equation}
is the \gls{FAS} response matrix, where each column $\vec{g}_l$ contains the antenna response for virtual port $l$. Notice that, in the conceptual \gls{FAS}, $\mat{G}$ is the identity matrix. 

Denoting by $\bm{\Sigma}_f^{{(\widetilde{m},m)}}$ the spatial channel covariance, i.e.,
\begin{equation}
    \bm{\Sigma}_f^{{(\widetilde{m},m)}} = \mathbb{E}\left[(\vec{f}^{(\widetilde{m},m)} - \mathbb{E}[\vec{f}^{(\widetilde{m},m)}])(\vec{f}^{(\widetilde{m},m)}- \mathbb{E}[\vec{f}^{(\widetilde{m},m)}])^H\right],
\end{equation}
the resulting \gls{FAS} covariance is simply 
\begin{align}
    \bm{\Sigma}_h^{{(\widetilde{m},m)}} =& \mathbb{E}\left[(\vec{h}^{(\widetilde{m},m)} - \mathbb{E}[\vec{h}^{(\widetilde{m},m)}])(\vec{h}^{(\widetilde{m},m)}- \mathbb{E}[\vec{h}^{(\widetilde{m},m)}])^H\right] \notag \\
    =& \mat{G}^T\bm{\Sigma}_f^{(\widetilde{m},m)}\mat{G}^*. \label{eq:Sigma_h}
\end{align}

\begin{remark}
    From \eqref{eq:Sigma_h}, it is clear that $\text{rank}(\bm{\Sigma}_h^{{(\widetilde{m},m)}}) \leq \text{min}\{\text{rank}(\bm{\Sigma}_f^{{(\widetilde{m},m)}}), \text{rank}(\mat{G})\}$. Hence, regardless the design of $\mat{G}$, the performance limit imposed by the spatial correlation cannot be circumvented, since the number of relevant eigenvalues of $\bm{\Sigma}_f^{{(\widetilde{m},m)}}$ is dictated by the physical aperture of the reconfigurable antenna \cite{Ramirez2024}. 
\end{remark}

Regarding communications performance, and assuming block fading, user $m$ will select the virtual port (antenna configuration) $\widehat{l}_m$ which maximizes its \gls{SINR}, i.e., 
\begin{align}
    \widehat{l}_m = \mathop{\text{arg max}}_{l} \gamma_m, \quad \gamma_m = \frac{\left|\vec{g}_l^T\vec{f}^{(m,m)}\right|^2}{\sum_{\widetilde{m}\neq m} \left|\vec{g}_l^T\vec{f}^{(\widetilde{m},m)}\right|^2+\sigma_w^2}, \label{eq:SINR}
\end{align}
where\footnote{As usual in wireless communications, we assume $\mathbb{E}[w_l^{(m)}]=0$.} $\sigma_w^2 = \mathbb{E}[|w_l^{(m)}|^2]$ $\forall$ $l$, leading to the normalized achievable rate
\begin{equation}
    R_m(\widehat{l}_m) = \log_2(1+\gamma_m(\widehat{l}_m)). \label{eq:Rate}
\end{equation}

\subsection{Metasurface-based fluid antenna}

\begin{figure}
    \centering
\includegraphics[width=1\linewidth]{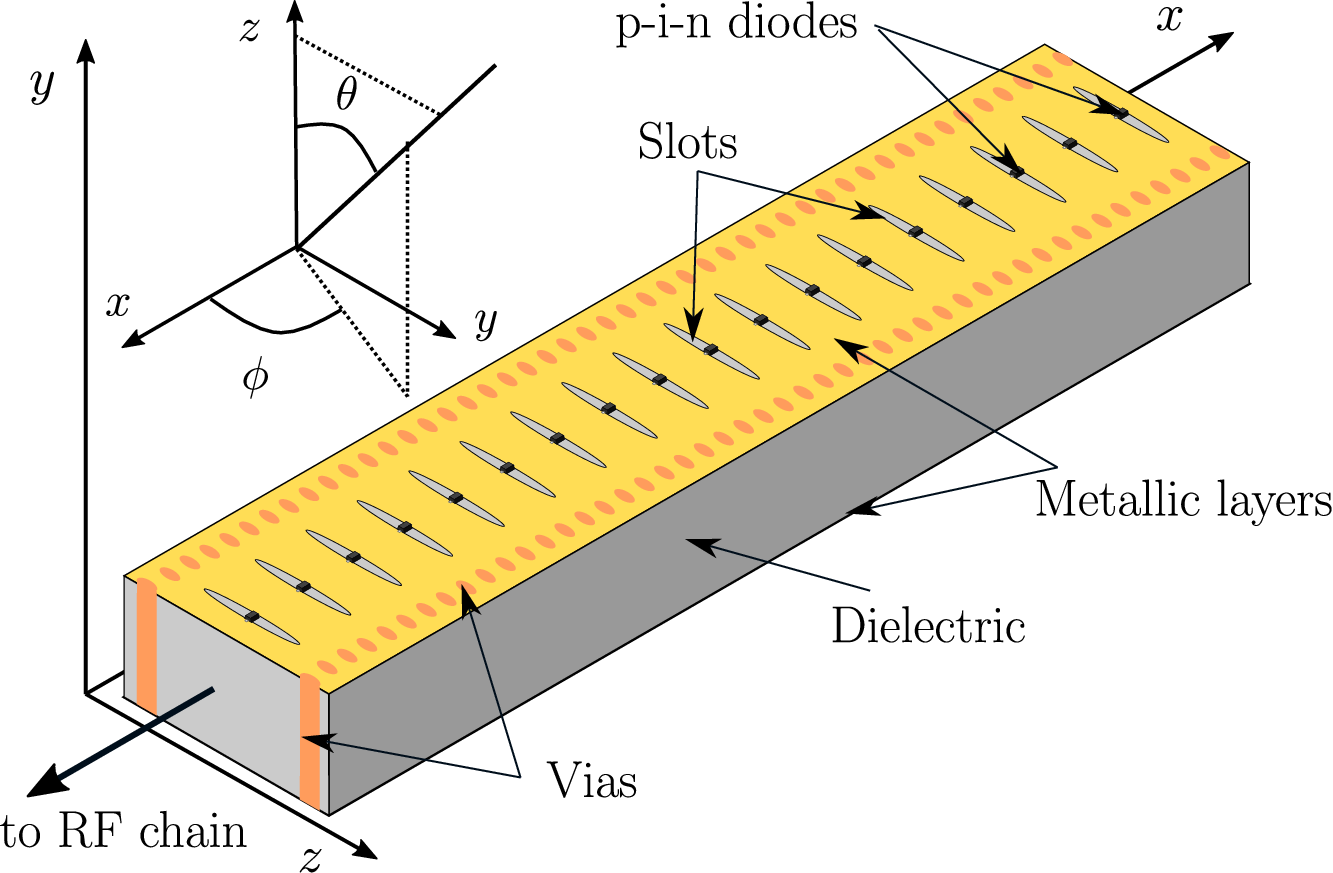}
    \caption{Schematic of metasurface-based FAS implementation in SIW and controlled by p-i-n diodes.}
    \label{fig:schematic}
\end{figure}

As an exemplary implementation of electronically reconfigurable \gls{FAS}, we adopt the metasurface-based design proposed in \cite{Ramirez2025}. Metasurfaces are cheap and scalable radiating structures that gained popularity as array replacement in \gls{MIMO} communications \cite{Schlezinger2021, Williams2022}. As illustrated in Fig. \ref{fig:schematic}, we consider a one-dimensional metasurface composed of a \gls{SIW}, which behaves as a rectangular waveguide from an electromagnetic viewpoint. The waveguide, of width $a$ ($z$-axis), height $b$ ($y$ axis) and length $c$ ($x$-axis), chosen such that only the fundamental mode propagates, is fed by a single \gls{RF} chain. On its upper face, $N$ closely-spaced slots act as radiating elements, behaving as magnetic dipoles. Reconfigurability is achieved by attaching a p-i-n diode to each slot, such that the slot can switch between two states: \textit{i)} if the diode is in direct polarization, then current flows through it effectively short-circuiting the slot (element `off'), \text{ii)} if the diode is in reverse polarization, in turn, the slot radiates (element `on'). By `on/off' switching of the different slots, the radiation pattern is modified. 

For the considered design, the responses $\vec{g}_l$ are given by\footnote{Note we have added the extra term $\real{Y_\text{rf}}^{1/2}/\sqrt{2}$ such that $|h_l^{(\widetilde{m},m)}x_{\widetilde{m}}|^2$ equals the received power from transmitting antenna $\widetilde{m}$.} \cite[Eq. (40)]{Ramirez2025}
\begin{equation}
    \vec{g}_l^T = \frac{1}{\sqrt{2}}\frac{\real{Y_\text{rf}}^{1/2}}{Y_\text{rf} + Y_{\text{in}}(\mat{Y}_\text{s}^{(l)})}\vec{y}_\text{w}^T(\mat{Y}_\text{s}^{(l)}+\mat{Y}_\text{c})^{-1}, \label{eq:gl_dma}
\end{equation}
where $Y_\text{rf}\in\mathbb{C}$ is the intrinsic parallel admittance of the \gls{RF} chain (receiver), $\vec{y}_\text{w}\in\mathbb{C}^{N\times 1}$ is the mutual admittance characterizing the propagation through the \gls{SIW} waveguide, $\mat{Y}_\text{C}\in\mathbb{C}^{N\times N}$ is the mutual coupling matrix, $Y_\text{in}(\cdot)\in\mathbb{C}$ is the input admittance of the metasurface, and 
\begin{equation}
    \mat{Y}_\text{s}^{(l)} = \text{diag}\left(\begin{pmatrix}
    y_{s1}^{(l)} & \dots & y_{sn}^{(l)} & \dots & y_{sN}^{(l)}
\end{pmatrix}\right)
\end{equation}
contains the load admittance of each slot. If slot $n$ is `on' (diode in reverse) for configuration $l$, then $y_{sn}^{(l)} = Y_\text{rad}$, being $Y_\text{rad}$ an arbitrary admittance characterizing the radiation properties of the slot. On the other hand, if slot $n$ is `off' (diode in direct), then $y_{sn}^{(l)} \rightarrow \infty$, which is equivalent to removing the corresponding row and column of $\vec{y}_\text{w}, \mat{Y}_\text{s}^{(l)}$ and $\mat{Y}_\text{c}$. The expressions for the different matrices in \eqref{eq:gl_dma} are functions of the metasurface structure, slot positions and employed dielectric, and the reader is gently referred to \cite[Sec. II]{Ramirez2025} or the provided code for further details. 

Importantly, notice that $Y_\text{in}$ depends on $\mat{Y}_\text{s}^{(l)}$. Therefore, each configuration yields a different input admittance seen by the \gls{RF} chain, hence varying the reflection coefficient and the total receiver power. Specifically, the reflection coefficient is given by \cite{Zhan2020}
\begin{equation}
    \Gamma^{(l)} = \frac{Y_\text{in}^*(\mat{Y}_\text{s}^{(l)}) - Y_\text{rf}}{Y_\text{in}(\mat{Y}_\text{s}^{(l)}) + Y_\text{rf}}. \label{eq:Gamma}
\end{equation}

\subsection{Wireless channel model}

From \cite[Sec. V-B]{Williams2022}, a general channel model is
\begin{equation}
    \vec{f} = \sqrt{\frac{\Omega_f K}{1+K}}\vec{a}(\theta_0, \phi_0) + \sqrt{\frac{\Omega_f}{1+K}}\vec{f}_\text{nlos}, \label{eq:Channel}
\end{equation}
where the dependence on the super-index $(\widetilde{m},m)$ is dropped for notational clarity, $\Omega_f = \mathbb{E}[\vec{f}^H\vec{f}]/N$, $K$ is the Rice factor\footnote{$K$ is defined as the ratio between the \gls{LoS} and \gls{NLoS} power.}, $\vec{f}_\text{nlos}$ is the \gls{NLoS} component, $\theta_0$ and $\phi_0$ are the polar and azimuth angle between the user and the transmitter, and 
\begin{equation}
    \vec{a}(\theta,\phi) = \begin{pmatrix}e^{i\vec{k}(\theta,\phi)^T\vec{r}_1} & e^{i\vec{k}(\theta,\phi)^T\vec{r}_2} & \dots & e^{i\vec{k}(\theta,\phi)^T\vec{r}_N}\end{pmatrix}^T
\end{equation}
with $\vec{k}(\theta,\phi) = k_0\begin{pmatrix}\sin(\theta)\cos(\phi) & \sin(\theta)\sin(\phi) & \cos(\theta)\end{pmatrix}^T$, $k_0$ the vacuum wavenumber, and $\vec{r}_n\in\mathbb{R}^{3\times 1}$ the cartesian coordinates of slot $n$. Assuming a rich scattering environment, $\vec{f}_\text{nlos}\sim\mathcal{CN}_N(\vec{0}, \mat{R}_\text{nlos})$, where the entries of the spatial correlation matrix are given by
\begin{equation}
    (\mat{R}_\text{nlos})_{n,n'} = \frac{\mathbb{E}_{\theta,\phi}[|G(\theta,\phi)|^2e^{i\vec{k}(\theta,\phi)(\vec{r}_n - \vec{r}_{n'})}]}{\mathbb{E}_{\theta,\phi}[|G(\theta,\phi)|^2]},
\end{equation}
where $G(\theta,\phi)$ is the farfield radiation pattern of the metasurface radiating elements (slots). The slots in Fig. \ref{fig:schematic} are modeled as infinitesimal magnetic dipoles, and hence $G(\theta,\phi) = \sqrt{3/2}\sin(\theta)$, leading to 
\begin{equation}
    (\mat{R}_\text{nlos})_{n,n'} = \frac{\mathbb{E}_{\theta,\phi}[\sin(\theta)^2e^{i\vec{k}(\theta,\phi)(\vec{r}_n - \vec{r}_{n'})}]}{\mathbb{E}_{\theta}[\sin(\theta)^2]}. \label{eq:SpatialCorrelation}
\end{equation}

\section{Conceptual vs reconfigurable FASs}
\label{sec:comparison}
Although reconfigurable antennas (like the considered metasurface) emulate the concept of \gls{FAS}, the different working principle---sampling space vs beamspace---has noticeable consequences. While conceptual \glspl{FAS} are limited \textit{only} by spatial propagation, electronically reconfigurable \glspl{FAS} are also limited by the beamforming capabilities of the antenna. This makes them lie in between conceptual fluid antennas and conventional \gls{MIMO}, inheriting antenna and codebook design from the latter. Besides, any practical implementation has to deal with reflections and impedance (admittance) matching, i.e., keeping \eqref{eq:Gamma} under control \cite{Zhang2024}. This translates into varying received power depending on the selected port; something largely ignored in the conventional formulation.

Regarding performance, we observe that the fundamental problem in \eqref{eq:SINR} is also different. In conceptual \glspl{FAS}, $\vec{g}_l$ are just canonical vectors, with their role being restricted to selecting one element of $\vec{f}$ (port). In turn, for reconfigurable \glspl{FAS}, \eqref{eq:SINR} resembles analog beamforming. Specifically, consider as an example that the system is dominated by interference, i.e., $\sigma_w^2\rightarrow 0$, which is the operational regime usually considered in FAMA. Then, the \gls{SINR} becomes
\begin{equation}
    \gamma_m\rvert_{\sigma_w^2\rightarrow 0} = \frac{\vec{g}_l^T\vec{f}^{(m,m)}\vec{f}^{(m,m)H}\vec{g}_l^*}{\vec{g}_l^T\left(\sum_{\widetilde{m}\neq m}\vec{f}^{(\widetilde{m},m)}\vec{f}^{(\widetilde{m},m)H}\right)\vec{g}_l^*},
\end{equation}
which is a special case of generalized Rayleigh quotient and it is optimized through a generalized eigendecomposition \cite{Ghojogh2023}. Interestingly, the interference matrix $\mat{B} = \sum_{\widetilde{m}\neq m}\vec{f}^{(\widetilde{m},m)}\vec{f}^{(\widetilde{m},m)H}$ is, at most, of rank $M-1$. Then, as long as $M<N$---always the case in \glspl{FAS}---$\mat{B}$ is rank deficient and, theoretically, it is possible to design $\vec{g}_l$ such that it aligns with the null space of $\mat{B}$. In other words, $\vec{g}_l^T\mat{B}\vec{g}_l^*\rightarrow 0$ and hence $\gamma_m\rightarrow \infty$. Notice that this is exactly the same working principle as in interference alignment \cite{ElAyach2012}.

Naturally, the optimal $\vec{g}_l$ is generally not achievable in practice since: \textit{i)} we are constrained to the beamspace of the reconfigurable antenna, and the closest solution would be projecting $\vec{g}_l$ onto the feasible space, \textit{ii)} perfect channel knowledge at each antenna element is required, which is unrealistic in \glspl{FAS} under a practical channel estimation and reconstruction protocol, and \textit{iii)} noise is always present, regularizing the denominator of $\gamma_m$. Nevertheless, the above observation suggests that electronically reconfigurable \glspl{FAS} might outperform conceptual \glspl{FAS}, potentially surpassing the performance limits predicted in the related literature. 

\section{Design of configurations for metasurface FAS}
\label{sec:codebook}
As discussed in the previous section, the design of the different achievable configurations is crucial to the performance of \glspl{FAS}, as a poor design can prevent the metasurface from fully exploiting its potential. Just as conceptual \glspl{FAS} densely sample the whole physical aperture, the idea here is to efficiently sample the whole beamspace provided by the metasurface. Notice that the problem reduces to \gls{MIMO} codebook design, but now the beamspace is constrained by the metasurface beamforming capabilities. Inspired by Grassmannian codebooks \cite{Love2003}, we propose a simple greedy method based on farthest configuration selection.

In our case, the beamspace is formed by all possible patterns realizable by the metasurface structure ---i.e., all the combinations of `on/off' elements. In general, though, it is desired that all the configurations share the same number of active elements such that the radiation efficiency remains approximately constant. This means that, for $N_\text{on}$ active elements, there are $\binom{N}{N_\text{on}}$ possible patterns. Therefore, an exhaustive search might be prohibitive for large $N$. The natural approach is to randomly sample the beamspace by choosing $C>L$ candidate configurations. Moreover, a reflection coefficient constraint is imposed for all the candidate configurations, such that the averaged received power does not vary significantly between ports, i.e., $|\Gamma_c| \leq \Gamma_{\rm th}$. 

Formally, we first generate a candidate codebook $\mat{C}\in\mathbb{C}^{P\times C}$ with $C$ configurations where each $\vec{g}_c$ in \eqref{eq:gl_dma} meets the reflection coefficient constraint. Denoting by $\vec{b}_c$ the configuration yielding the response $\vec{g}_c$, then each column of $\mat{C}$ stores the farfield pattern generated by $\vec{b}_c$ and calculated over a sphere of radius $q\gg 1$, which is computed as \cite[Eq. (33)]{Ramirez2025}
\begin{equation}
    H_{\rm ff}(\vec{r})=\sum_{n=1}^Ne^{-ik_0\|\vec{r}-\vec{r}_n\|_2}\left(\vec{g}_c\right)_n
\end{equation}
where $\vec{r} = q\begin{pmatrix}\cos(\phi_p) & \sin(\phi_p) & 0\end{pmatrix}^T$. Here, $\phi_p\in[0,\pi)$ for $p=1,\dots P$ are equally spaced azimuth\footnote{Since a linear metasurface is considered, thus lacking beamforming capabilities in elevation angles, we restrict our codebook design to azimuth directions. However, this can be trivially extended to elevation and azimuth for planar implementations.} directions measured from the metasurface center. 

Once $\mat{C}$ is generated and we have the candidate set $\{\vec{b}\}_{c=1}^C$, we define $\widehat{\mat{C}}$ as the column-wise normalized version, i.e., column $\widehat{\vec{c}}_c$ of $\widehat{\mat{C}}$ is obtained from column ${\vec{c}}_c$ of ${\mat{C}}$ as $\widehat{\vec{c}}_c = {\vec{c}}_c/\|{\vec{c}}_c\|_2$. With the normalized codebook, we define a distance metric between candidate configurations $\vec{b}_c$ and $\vec{b}_{t}$ as $d_{c,t} = |\widehat{\vec{c}}_c^H\widehat{\vec{c}}_t|$, which represents a measure of the pattern overlap. This metric can be seen as the equivalent of chordal distance in Grassmannian codebooks. 

For the final codebook $\{\vec{b}\}_{l=1}^L$, we choose the first configuration $\vec{b}_1$ either randomly or based on any other criterion, such as maximum norm $\|\vec{c}_l\|_2$. For the remaining codewords, configuration $l+1$ is chosen such that it minimizes the maximum correlation to the set $\{\widehat{\vec{c}}\}_{c=1}^l$. In other words, for $l = 2, \dots, L$, we solve
\begin{equation}
    \vec{b}_{l+1} = \mathop{\text{arg min}}_c \mathop{\text{max}}_{t\leq l} d_{t,c}.
\end{equation}
\begin{remark}
    We have imposed that each configuration yields $|\Gamma_l| \leq \Gamma_{\rm th}$. However, $\Gamma_l$ directly depends on the \gls{RF} chain admittance $Y_\text{rf}$ . To obtain $\Gamma_l = 0$, then $Y_\text{rf} = Y_\text{in}^*(\mat{Y}_\text{s}^{(l)}) $\eqref{eq:Gamma}, which means that for fixed $Y_\text{rf}$ one cannot match all the configurations. Ideally, one might add a reconfigurable matching network, but this implies additional hardware and control logic. \Glspl{FAS} are supposed to be simple and efficient, and hence we stick to fixed $Y_\text{rf}$ (equivalently, fixed matching network), which is chosen such that low reflection coefficients are obtained in as many configurations as possible. Specifically, for a set $\{\vec{b}_t\}_{t=1}^T$ of randomly generated configurations and corresponding input admittances, we select $Y_\text{rf}$ as the geometric median of $\{Y_\text{in}(\mat{Y}_\text{s}^{(t)})\}_{t=1}^T$, i.e, $Y_\text{rf} = \text{arg min}_X \sum_t |Y_\text{in}(\mat{Y}_\text{s}^{(t)})-X|$ \cite{Bose2003}.
\end{remark}

\section{Performance evaluation of metasurface-based FAMA}
\label{sec:performance}
We now compare the performance of electronically reconfigurable and conceptual \glspl{FAS}, aiming to validate the rationale in Section \ref{sec:comparison}. To that end, we focus on a generic \gls{FAMA} system with signal model as in \eqref{eq:FAS_signal}, and we use the sum-rate $R_\text{sum} = \sum_m^M R_m$ with $R_m$ in \eqref{eq:Rate} as performance metric. Importantly, perfect channel knowledge at every port is assumed, i.e., $\vec{f}^{(\widetilde{m},m)}$ is known for every $\widetilde{m},m$. Therefore, the following results and discussion can be seen as the architectural limits of each system. 

For all the simulations, we assume a \gls{SIW} waveguide of dimensions (in millimeters) $a = 58$, $b = 23.2$ and $c = 25*(N-1)+60$. The dielectric has a relative permittivity of $3.55$, and the spacing between slots is $25$ mm, corresponding roughly to $\lambda_0/5$ ($\lambda_0$ denotes wavelength in the air) at a frequency of $2.4$ GHz. The slots are characterized by $Y_\text{rad} = -i0.0037$ for a dipole normalized length $l_m = 0.0019$ (see \cite[Remark 1]{Ramirez2025}). The wireless channel follows \eqref{eq:Channel}, and the channel spatial correlation is given by \eqref{eq:SpatialCorrelation} with fixed $\theta = \pi/2$ (2D propagation environment) and azimuth distribution 
\begin{equation}
    f_\phi(\phi; \alpha) = \frac{\Gamma(1+\alpha/2)}{\sqrt{\pi}\Gamma((1+\alpha)/2)}\sin^\alpha(\phi),
\end{equation}
where $\alpha\geq 0$ is a shape parameter and $\Gamma(\cdot)$ is the Gamma function. The larger $\alpha$, the more concentrated the arrival angles around broadside ($\phi = \pi/2$). Notice that, if $\alpha=0$, then $f_\phi(\phi; \alpha)$ represents the uniform distribution and \eqref{eq:SpatialCorrelation} reduces to Jakes' correlation. 

The noise power $\sigma_w^2$ is defined based on a target \gls{SNR} $\rho$ in the single user case. That is, 
\begin{align}
    \rho_\text{cpt} &= \frac{\Omega_f}{\sigma_{w,\text{cpt}}^2} \;\;\text{(conceptual FAS)}, \\
    \rho_\text{meta} &= \frac{\mathbb{E}_L[\vec{g}_l^H\vec{g}_l]}{\sigma_{w,\text{meta}}^2} \;\;\text{(metasurface-based FAS)},
\end{align}
such that we keep the ratio average received power to noise in both systems. For additional simulation details that might be needed to reproduce the results, readers are referred to the provided code. 

\begin{figure}[t]
    \centering
    \includegraphics[scale=0.5]{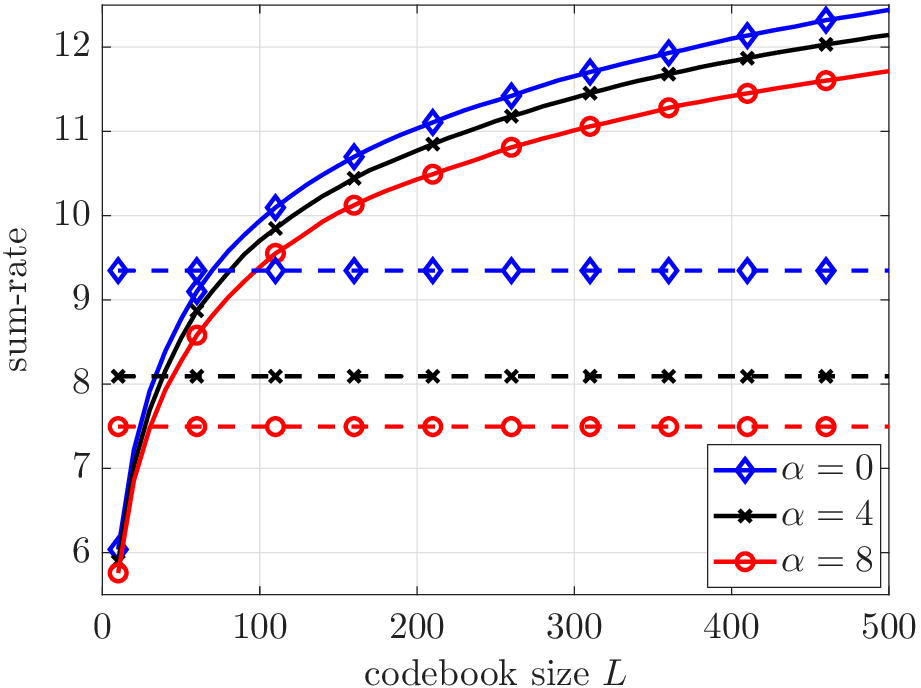}
    \caption{Sum-rate vs codebook size and spatial correlation for a metasurface-based FAS with $N=24$ and $N_\text{on}=14$ (solid lines) and a conceptual FAS (dashed lines). $M = 3$, $K = 0$, $\rho = 20$ dB and $C = 10^4$.}
    \label{fig:figure_1}
\end{figure}

With the above definitions, we begin exploring the impact of codebook size $L$ (number of virtual ports) and of spatial correlation, considered the limiting factor in conceptual \glspl{FAS}. Different codebooks are generated according to the method described in Section \ref{sec:codebook} for a metasurface of $N = 24$ elements with $N_\text{on} =  14$ active elements, yielding a total physical aperture of $5\lambda_0$. For the conceptual \gls{FAS}, the same physical aperture is used, with 40 equally spaced ports (sampling positions) per $\lambda_0$; a number that effectively saturates the performance \cite{Ramirez2024}. The comparison is shown in Fig. \ref{fig:figure_1} for $M = 3$ users, $\rho = 20$ dB (interference-limited) and $K =0$ (pure \gls{NLoS}). Interestingly, we confirm that the intuition in Section \ref{sec:comparison} is aligned with the results, and electronically reconfigurable \glspl{FAS} significantly outperform the conceptual one (for the same physical aperture), regardless the angular distribution and hence correlation structure. Two observations can be made: \textit{i)} the impact of spatial correlation is much lower for the metasurface implementation, which confirms that a main benefit is made from interference cancellation via eigenvector projection, and \textit{ii)} metasurfaces can exploit multiple virtual ports (radiation patterns), rendering a richer effective beamspace (and thus higher chances of projecting closer to the null-space). 

\begin{figure}[t]
    \centering
    \includegraphics[scale=0.5]{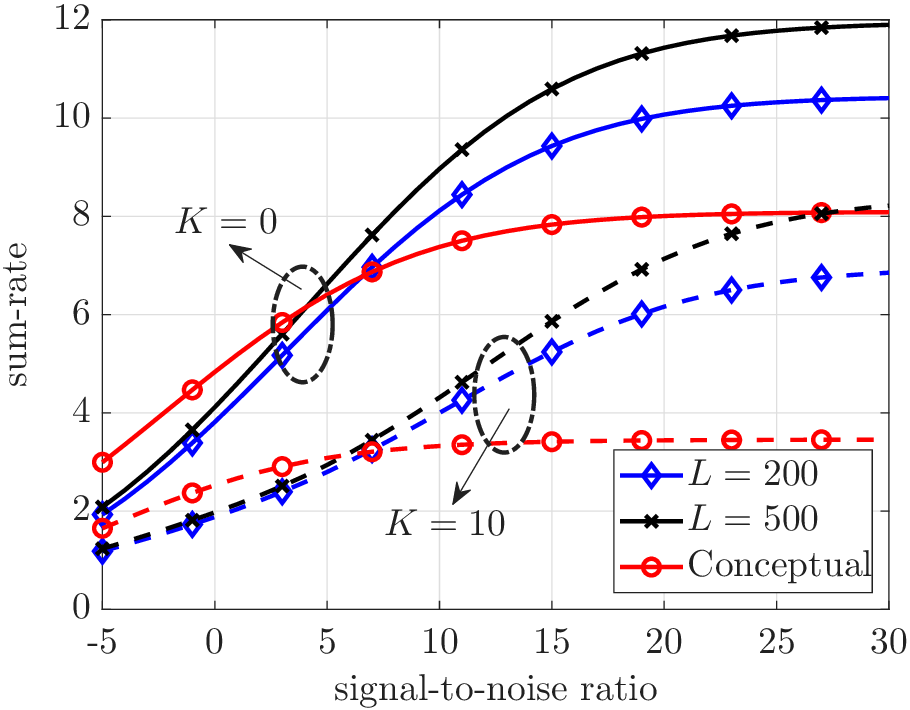}
    \caption{Sum-rate vs SNR $\rho$ and Rice factor $K$ for a metasurface-based FAS with $N=24$ and $N_\text{on}=14$ and a conceptual FAS. $M = 4$, $\alpha = 0$ and $C = 10^4$.}
    \label{fig:figure_2}
\end{figure}

The same setup is used to evaluate the performance against noise power. To that end, Fig. \ref{fig:figure_2} plots the sum-rate for the same metasurface-based and conceptual \glspl{FAS} for varying $\rho$ (noise power) and different $K$ factors. $M = 4$ users are considered and Jakes' correlation ($\alpha=0$) is assumed, as it is the widely-used baseline benchmark in the fluid antenna literature. We first observe that, as expected, a stronger \gls{LoS} reduces diversity and thus multipath gains. This is also a well-known fact in conventional \gls{MIMO}, where rich spatial scattering is beneficial. Regarding the impact of noise, we see again that for large $\rho$ (interference-limited) the metasurface-based \gls{FAS} yields much better results (up to $50\%$ sum-rate increase). However, as noise power increases (i.e., lower $\rho$), the denominator in \eqref{eq:SINR} becomes dominated by noise and canceling out interference yields diminishing returns. In fact, if noise dominates, conceptual \gls{FAS} performs better. 

\begin{figure}[t]
    \centering
    \includegraphics[scale=0.5]{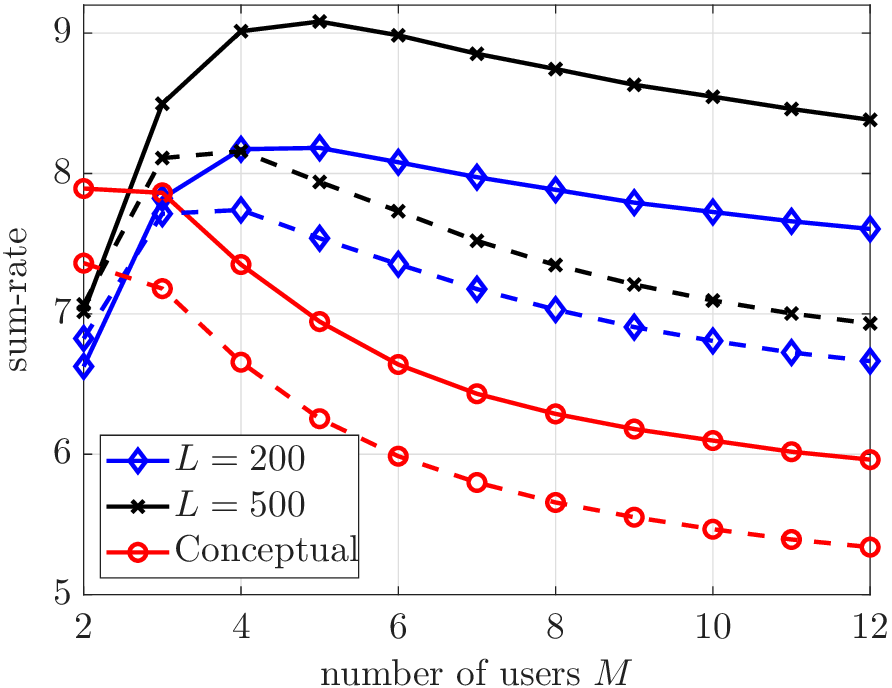}
    \caption{Sum-rate vs number of users for a metasurface-based FAS with $N=24$ and $N_\text{on}=14$ (solid lines) and $N=16$ and $N_\text{on}=8$ (dashed lines) and conceptual FASs of the same physical apertures. $K = 0$, $\alpha = 0$, $\rho = 10$ dB, and $C = 10^4$ for $N=24$ and $C=700$ for $N=16$.}
    \label{fig:figure_3}
\end{figure}

Finally, we evaluate both systems in terms of the number of simultaneous users $M$. Together with the so far considered metasurface, we also evaluate the performance of a smaller one with $N=16$ and $N_\text{on} = 8$ ($3.5\lambda_0$ of physical aperture). The results are sown in Fig. \ref{fig:figure_3}, where $K=0$, $\alpha = 0$, and $\rho = 10$ dB (noise does not dominate but still has an impact). Importantly, electronically-reconfigurable \glspl{FAS} support a larger number of users without excessively degrading the performance. Even though the sum-rate starts decreasing for more than $4$ or $5$ users (depending on the physical aperture), which indicates a degradation in the individual rates, the number of supported users is considerably larger than what conceptual \glspl{FAS} (and then state-of-the-art works) predict. 

\section{Conclusions and open questions}
\label{sec:Conclusions}
This paper evaluates the performance of metasurface-based \glspl{FAS}, and compares them with the conceptual idea of \gls{FAS}, where a single antenna element can be freely moved within a predefined aperture. The results and discussions presented here suggest that we should, at the very least, question the performance limits and conclusions drawn in the \glspl{FAS} literature. First, spatial correlation does not seem to be the only limiting factor, since reconfigurable \glspl{FAS} maximizes the received \gls{SINR} by canceling interference through projecting the metasurface response onto the interference null-space, leading to maximum results when the system is interference-limited. This mechanism is different from the one employed in conceptual \glspl{FAS}, where we opportunistically seek for a spatial point where interference is at a minimum. The larger the number of simultaneous users, the more noticeable the performance gap.

The previous point leads to the follow-up question: \textit{how to design the different metasurface configurations?}, which resembles to the \gls{MIMO} codebook design problem. Hence, practical implementations of \glspl{FAS} stand at a crossroads between classical \gls{MIMO} and conceptual \glspl{FAS}. In this work, we propose a general codebook design that aims to explore the whole feasible beamspace while keeping the reflection coefficient (and then antenna radiation efficiency) under control. The chosen criterion is correlation between radiated field patterns, but other methods should be explored, considering maybe procedures that account for spatial correlation characteristics. 

Last, in conceptual \glspl{FAS} the number of required ports to achieve the maximum performance is dictated by spatial correlation and, consequently, by the physical aperture. For electronically reconfigurable \glspl{FAS}, a larger number of ports may be required to effectively sample the feasible beamspace, which in turns complicates channel estimation. Beamspace correlation can be exploited to reduce the number of estimations, but specific algorithms are necessary and the impact of this overhead need to be quantified.

\bibliographystyle{IEEEtran}
\bibliography{references.bib}

\end{document}